\documentclass[twocolumn,superscriptaddress,showpacs,aps,prd,nofootinbib,amsmath,amssymb,nobibnotes,floatfix]{revtex4-1}

\usepackage{hyperref}
\usepackage{amsfonts}
\usepackage{amsmath}
\usepackage{mathrsfs}
\usepackage{epsfig}
\usepackage{graphicx}               
\usepackage{url}
\usepackage{hyperref}
\usepackage{float}
\usepackage{color}
\usepackage[utf8]{inputenc}

\usepackage{graphicx}
\usepackage{epsf}

\usepackage{comment}

\newcommand{\nc}{\newcommand}

\nc{\beq}{\begin{equation}}
\nc{\eeq}{\end{equation}}
\nc{\beqa}{\begin{eqnarray}}
\nc{\eeqa}{\end{eqnarray}}

\usepackage{slashed}

\newcommand{\lsim}{\!\mathrel{\hbox{\rlap{\lower.55ex \hbox{$\sim$}} \kern-.34em \raise.4ex \hbox{$<$}}}}
\newcommand{\gsim}{\!\mathrel{\hbox{\rlap{\lower.55ex \hbox{$\sim$}} \kern-.34em \raise.4ex \hbox{$>$}}}}

\def\be{\begin{equation}}
\def\ee{\end{equation}}

\newcommand\affspc{\vspace{4pt}}

\usepackage{footmisc}
\usepackage{setspace}

\begin{document}

\title{Magnetosphere of a spinning black hole and the role of the current sheet}

\author{William E.\ East}
\affiliation{Perimeter Institute for Theoretical Physics, Waterloo, Ontario N2L 2Y5, Canada \affspc}
\author{ Huan Yang}
\affiliation{University of Guelph, Guelph, Ontario N2L 3G1, Canada}
\affiliation{Perimeter Institute for Theoretical Physics, Waterloo, Ontario N2L 2Y5, Canada \affspc}

\begin{abstract}
We revisit the problem of a spinning black hole immersed in a uniformly
magnetized plasma within the context of force-free electrodynamics. Such
configurations have been found to relax to stationary jetlike solutions
that are powered by the rotational energy of the black hole. We write down
an analytic description for the jet solutions in the low black hole spin
limit, and demonstrate that it provides a good approximation to the
configurations found dynamically. For characterizing the magnetospheres of
rapidly spinning black holes, we find that the current sheet which forms in
the black hole ergosphere plays an essential role. We study the properties of the
current sheet, and its importance in determining the jet solution and the
rate at which energy is extracted from the black hole. 
\end{abstract}

\maketitle

\section{Introduction}
From active galactic nuclei to ultraluminous x-ray binaries to short gamma-ray
bursts, many of the most powerful astrophysical sources of electromagnetic
radiation are thought to be associated with the magnetospheres of black holes
(BHs).  
A central paradigm often invoked in explaining these observations is the
Blandford-Znajek mechanism~\cite{Blandford:1977ds}, whereby a strongly
magnetized plasma can tap into a spinning BH's reservoir of rotational energy
to power a jetted outflow of energy.  In general, the description
of an accreting BH will involve complicated fluid dynamics,
radiation, thermal effects, and so on. However, in many cases the essential effects can be captured
using force-free electrodynamics (FFE) which is appropriate for a tenuous 
plasma nearby the BH, where the magnetic energy density dominates over the 
matter density and pressure.
In this description, only the dynamics of the electromagnetic fields needs to 
be kept track of, with the matter assumed to be arranged by the strong magnetic
field in such a way that the Lorentz force always vanishes.

Here we study a prototypical setup that exhibits the Blandford-Znajek
mechanism: a spinning BH immersed in a uniformly magnetized plasma.
Starting with~\cite{Komissarov:2004ms}, there have been a number of studies
of this problem using numerical evolutions of the general-relativistic equations of 
FFE~\cite{Komissarov:2007rc,Palenzuela:2010xn,Paschalidis:2013gma,Yang:2015ata,Carrasco:2017ucd}. 
These have found that the solution relaxes to a stationary BH jet 
configuration with a Poynting flux powered by the rotational energy of
the BH.
This provides a simple setting to study the underlying mechanisms of relativistic 
BH jets. 

In~\cite{Yang:2015ata}, a whole family of stationary BH jet solutions
describing a slowly spinning BH in a uniformly magnetized plasma was presented.
However, no sign of mode instability on the relevant time scales was found for
these solutions (see also \cite{yang2014stability} for a study of the stability
of Blandford-Znajek split-monopole type magnetosphere configurations), and it
was left as an open question what condition picked out the unique solution
found by the numerical evolutions. 

There have also been several
studies~\cite{Nathanail:2014aua,Pan:2015imp,Pan:2017npg} attacking this problem
by numerically solving the Grad-Shafranov equation which governs an
axisymmetric, stationary solution to the force-free equations. 
Such an approach requires the prescription of suitable boundary conditions,
both at infinity, and at any current sheets, which is not always
straightforward.

In this work, we revisit this problem guided by high accuracy solutions, again
obtained by evolving the general-relativistic force-free equations.  We confirm
previous results regarding the structure of the electromagnetic fields, the
angular velocity, and luminosity from these
solutions~\cite{Komissarov:2004ms,Palenzuela:2010xn,Paschalidis:2013gma,Yang:2015ata,Carrasco:2017ucd}.
However, we also focus on some new aspects.  For generic BH spins, we
explicitly demonstrate that these solutions obey a relation between the current
and angular velocity of magnetic field lines which can be seen as an outgoing
radiation condition at infinity.  In the small-spin regime, where the influence
of the current sheet turns out to be negligible, this condition can be used to
derive a unique jet solution. We demonstrate that this analytic solution
provides a good description of the luminosity and other properties of the jet
solutions at low, and even moderate spins.

At higher values of BH spin, we find that the current sheet that develops
within the BH ergosphere plays an important role.  As described
in~\cite{Komissarov:2004ms} (see also~\cite{Carrasco:2017ucd}), part of the
flux of energy and angular momentum from the jet can be traced to the current
sheet, as opposed to the BH horizon. 
We add to this previous work by quantifying this difference
as a function of spin, finding that
roughly half of the total power comes from the current sheet for rapidly
spinning BHs.  We also discuss the breakdown of the force-free approximation at the
current sheet due to the loss of magnetic dominance, and how different
treatments of this affect the field configurations.  We find that the magnitude
of the jet power is actually insensitive to the details of this treatment.

The remainder of this paper is organized as follows. In Sec.~\ref{sec:methods},
we describe the methods we use to evolve the equations of FFE 
and extract relevant quantities from the resulting BH jet solutions.
In Sec.~\ref{sec:low_spin}, we describe the relation between current, angular velocity, and
flux of magnetic field lines found in all the jet solutions, including how it is related to
an outgoing radiation condition, and how it can be used to derive an analytic solution
for slowly spinning BHs.
We present the results we find for BH jet solutions by evolving with
FFE in Sec.~\ref{sec:results}. We discuss these
and compare them to other results in the literature in Sec.~\ref{sec:discuss}.

\section{Methods}
\label{sec:methods}
The equations of FFE are just the Maxwell equations: $\nabla_a F^{ab}=J^b$
and $\nabla_{[ a}F_{bc ]}=0$, where $F^{ab}$ is the field strength tensor
and $J^b$ is the four current, supplemented by the force-free condition $F_{ab}J^b=0$. 
We discretize these equations using 
fourth-order Runge-Kutta time stepping and standard fourth-order stencils for
spatial derivatives, as described in~\cite{East:2015pea}. We evolve the FFE equations on a
fixed BH spacetime with mass $M$ and dimensionless spin $a$ using
Cartesian Kerr-Schild coordinates~\cite{1965cngg.conf..222K}.  We restrict
ourselves to axisymmetric configurations which we evolve using the modified
cartoon method introduced in~\cite{Pretorius:2004jg}, where we take our
numerical domain to be the two-dimensional half plane given by $0\leq x <
\infty$ and $-\infty < z < \infty$. Spatial infinity is included on the grid
through the use of compactified coordinates.
Derivatives in the $y$ direction are
calculated by rewriting them in terms of $x$ and $z$ derivatives using
axisymmetry, and regularity is imposed at the $z$ axis.  
As noted in~\cite{Yang:2015ata}, evolutions of the same configurations considered
here that do not enforce axisymmetry still find an axisymmetric relaxed state.

We use six levels of
mesh refinement, with 2:1 refinement ratio to concentrate resolution around
the BH.  The finest level has a resolution of least $dx\approx 0.02M$,
though for some cases we use up to 4 times higher resolution.  The use of
axisymmetry and mesh refinement allows us to evolve these configurations for
long times, to ensure that they have fully relaxed towards stationary
solutions, and also have sufficient resolution to capture the effects of small
and near extremal BH spins.

For the electromagnetic fields we use a uniform magnetic field as the
initial condition, setting  $E^i=0$ and $B^i=\delta^i_zB_0/\sqrt{\gamma}$. 
The factor of the determinant of the spatial metric $\gamma$ is included so that
the magnetic field will be divergenceless on the BH spacetime.
At spatial infinity, we leave the field fixed during the evolution, which is causally disconnected from the solution in the interior where we measure all the relevant quantities.

Here and throughout we use Lorentz-Heavyside units with
$G=c=1$.

\subsection{Treatment of loss of magnetic dominance}
\label{ss:mag_dom}
A generic occurrence when evolving the FFE equations
is the development of regions where magnetic dominance is lost, i.e.
where $B^2<E^2$, or equivalently in terms of the field strength tensor
$F^2:=F^{ab}F_{ab}=2(B^2-E^2)<0$. When this occurs, the FFE equations are no longer
hyperbolic~\cite{Komissarov:2002my,Palenzuela:2011es,Pfeiffer:2013wza}, and some 
\emph{ad hoc} prescription must be applied.
Here we adopt a common prescription~\cite{Komissarov:2004ms,Spitkovsky:2006np,Palenzuela:2010xn}
and reduce the magnitude
of the electric field so that it no longer 
exceeds that of the magnetic field: 
\beq
E^i\rightarrow E^i\times(B^2/E^2)^{1/2} \ .
\eeq
This acts as a source of dissipation which---though artificial---mimics
the loss of electromagnetic energy due to a strong electric field accelerating
particles.  In some circumstances this has been found to provide good agreement
with the electromagnetic dissipation seen in kinetic
simulations~\cite{East:2015pea,Zrake:2015hda,Nalewajko:2016jum,Yuan:2016fyz}, though of
course only the kinetic calculation captures the resulting particle
acceleration.

In the cases studied here, loss of magnetic dominance only occurs at the
current sheet which forms in the equatorial plane of the BH ergosphere.  We
discuss this in detail below, and consider how different prescriptions for
treating this region affects the resulting solution.

\subsection{Measured quantities}
\label{ss:measure}
As mentioned above, we use Cartesian Kerr-Schild coordinates,
and in this paper $\{x,y,z\}$ refer to these coordinates.
However, we will use $\theta$ to refer to the Boyer-Lindquist
polar angle.

An axisymmetric, stationary solution can be described in terms of a magnetic
flux function $\psi$, polar current $I$, and the angular velocity of fields
lines $\Omega_F$, where the latter two quantities are constant along magnetic
field lines, and hence are just functions of $\psi$.
Following~\cite{Gralla:2014yja}, one can define $\psi$
by integrating the field strength tensor along a surface $\mathcal{S}$ bounded by
a curve of revolution in the azimuthal direction:
\beq
\psi = \frac{1}{2 \pi} \int_{\mathcal{S}} F \ .
\eeq
The other two quantities can be computed
from the field strength tensor and its dual as $\Omega_F=F_{ab}t^a\theta^b/F_{ab}\theta^a\phi^b$ 
and $I=2\pi {}^*F_{ab}t^a\phi^b$, where $t^a$ and $\phi^a$ are the time and axisymmetric
Killing vectors, and $\theta^a$ points in the polar direction.

The flux of energy $\dot{\mathcal{E}}$ and angular momentum $\dot{\mathcal{J}}$ through a surface generated
by a polodial curve $\mathcal{P}$ can be written in terms of these quantities as 
\beq
\dot{\mathcal{E}} = -\int_{\mathcal{P}} \Omega_F I d\psi
\eeq
and 
\beq
\dot{\mathcal{J}} = -\int_{\mathcal{P}} I d\psi \ .
\eeq

We also compute two different measurements of the energy density in the 
electromagnetic fields.
The first is the energy density 
as seen by a set observers with four velocity $n_a$ perpendicular to slices
of constant coordinate time
\beq
\rho_{\rm EM}=n_a n_b T^{ab}=\frac{1}{2}(E^2+B^2) 
\eeq
where $T^{ab}$ is the electromagnetic stress-energy tensor.
The second measure of energy density is the one with respect to the 
timelike Killing vector of the Kerr spacetime, $\rho_K=n^a t^b T_{ab}$.
The volume integral of this quantity is conserved, modulo any flux through
the BH horizon, or the breakdown of FFE.
In contrast to $\rho_{\rm EM}$ which is always $\geq 0$,  $\rho_K$ can become negative within the BH
ergosphere.

\section{Radiation condition and the slowly spinning black hole jet solution}
\label{sec:low_spin}

In this section we motivate the relation between $I$, $\Omega_F$, and $\psi$
which we empirically find to hold in our BH jet solutions, and use this
relation to derive an analytic description of these jet solutions in the limit
of a slowly spinning BH.

\subsection{Outgoing radiation condition}
At distances much greater than the size of the BH, it is reasonable to assume
that the jet solution becomes translationally invariant along the z-axis. As a
result, $\psi$ is a function of the cylindrical radius $\rho := r \sin \theta$. 
Using an orthonormal basis: $\{\hat\rho,\hat\theta,\hat z \}$, 
the Grad-Shafranov equation is equivalent
to
 \begin{align}
 j_{\hat \phi} B_{\hat z}-j_{\hat z} B_{\hat \phi}+ q E_{\hat \rho} =0\,,
 \end{align}
with
\begin{align}
j_{\hat \phi} &= \frac{d}{d \rho} \left ( \frac{1}{\rho} \frac{d \psi}{d \rho}\right )\,, \quad j_{\hat z}= \frac{1}{2 \pi \rho}\frac{d I}{d \rho}\,, \nonumber \\
B_{\hat z} & = \frac{1}{\rho} \frac{d \psi}{d \rho}\,,\quad B_{\hat \phi} =\frac{I}{2 \pi \rho} \,,\nonumber \\
E_{\hat \rho} & =\rho \,\Omega_F B_z\,,\quad q = \frac{1}{\rho} \frac{d}{d \rho}\left ( \rho\, \Omega_F \frac{d \psi}{ d \rho}\right )\,.
\end{align}
Here $j^i$ is the three-current and $q$ is the charge density. 

Previous works \cite{Nathanail:2014aua,Pan:2015imp,penna2015black,Pan:2017npg} have discussed
the scenarios with $E_{\hat \rho}=\pm B_{\hat \phi}$ or ${\bf E} =\mp \hat{z} \times {\bf
B}$, which correspond to the ingoing and outgoing ``radiation condition,"
respectively. In such cases, $j_{\hat \phi}=0$, and $B_{\hat z}$ is constant within the jet
\footnote{Notice that an inconsistent expansion of the Grad-Shafranov equation
in spherical coordinates in \cite{Pan:2015imp} neglects the $j_{\hat \phi}$ term,
which should appear in the same expansion order, and consequently makes the
authors  claim a derivation of the ``radiation conditions".}, which implies
$I=\pm 4\pi\Omega_F \psi$. Reference~\cite{Brennan:2013kea} also discuss how
having an ingoing/outgoing dynamical wave 
implies that 
${\bf E} =\mp \hat{n} \times {\bf B}$, where $\hat{n}$
is the outgoing unit normal.
Strictly speaking, as the final jet solution is
stationary in time, it is less clear \emph{a priori} whether the requirements on its electromagnetic fields have
the same physical meaning as ingoing and outgoing conditions for dynamical
fields. We also note that due to the presence of the uniform magnetic field, these
solutions do not satisfy this outgoing wave relation outside the jet tube.

Nevertheless, as discussed below in Sec.~\ref{sec:results}, we find that
the stationary solutions that we obtain for generic BH spins indeed
satisfy $I=-4\pi\Omega_F\psi$.
This condition allows for the derivation of a unique jet solution in the limit of low BH spin,
as we will now outline.

\subsection{Low spin solution}
\label{ss:low_spin}
To derive the low spin solution, we use the fact that the spatial
dependence of the flux function is unchanged from the Schwarzschild
Wald-type solution~\cite{Wald:1974np}, to leading order~\cite{Blandford:1977ds}  
\beq\label{eqn:psi}
\psi = \frac{1}{2}B_0\rho^2
\eeq
where $\rho=r \sin \theta$ (in Boyer-Lindquist coordinates).
If we combine this with the Znajek condition that comes
from demanding regularity on the horizon, 
\beq\label{eqn:i}
I=2\pi(\Omega_F-\Omega_H)\left(\frac{d\psi}{d\theta}\right) \frac{(r_+^2+(aM)^2)\sin \theta}{r_+^2+a^2M^2\cos^2\theta}
\eeq
and the requirement
that $I=-4\pi\Omega_F \psi$, then we arrive at the relation on the horizon:
\beq
\Omega_F=\Omega_H\left(\frac{|\cos\theta|}{1+|\cos\theta|}\right) \ .
\label{eqn:omega_theta}
\eeq
Here $\Omega_H=\frac{a}{2r_+}$ and $r_+=M(1+\sqrt{1-a^2})$ are the BH horizon rotational
frequency and radius.
We then have that
\beq
\Omega_F=\Omega_H\left(\frac{\sqrt{1-\bar{\psi}}}{1+\sqrt{1-\bar{\psi}}}\right) \ ,
\label{eqn:omega_psi}
\eeq
where $\bar{\psi}:=\psi/(2B_0M^2)$, for all field lines crossing the BH horizon.
For all other field lines (i.e.\ those with $\bar{\psi}\geq1$) we
have that $I=\Omega_F=0$.
This is the same as the relation written down in~\cite{Beskin:2010iba,Pan:2014bja,Pan:2017npg}. 

From Eqs.~\eqref{eqn:psi},~\eqref{eqn:i}, and~\eqref{eqn:omega_psi}, we can calculate the flux of energy $\dot{\mathcal{E}}$ and
angular momentum $\dot{\mathcal{J}}$ in the low spin limit. They are given by 
\beq
\dot{\mathcal{E}}=128\pi\left(\frac{17}{24}-\log{2}\right)B_0^2M^4\Omega_H^2
\label{eqn:edot}
\eeq
and
\beq
\dot{\mathcal{J}}=\frac{16}{3}\pi B_0^2M^4\Omega_H.
\label{eqn:jdot}
\eeq

The field tensor in the slow rotation limit is then given by
\begin{align}
F  = \,& d \psi \wedge (d \phi - \Omega_F dt) +\frac{I}{2 \pi} \frac{dr \wedge d \theta}{f \sin\theta} \nonumber \\
= \,& B_0 \rho\, d \rho \wedge d \phi -\frac{B_0 \rho \,\Omega_H \sqrt{1-\bar{\rho}^2}}{1+\sqrt{1-\bar{\rho}^2}} d \rho \wedge dt \nonumber \\
& -\frac{B_0 \rho^2 \Omega_H}{f \sin\theta } \frac{\sqrt{1-\bar{\rho}^2}}{1+\sqrt{1-\bar{\rho}^2}} dr \wedge d \theta\,,
\end{align}
with $f := 1-2M/r$ and $\bar{\rho}$ defined to be $\rho/(2M)$. 

\section{Results}
\label{sec:results}
\begin{figure}
\begin{center}
\includegraphics[width=\columnwidth,draft=false]{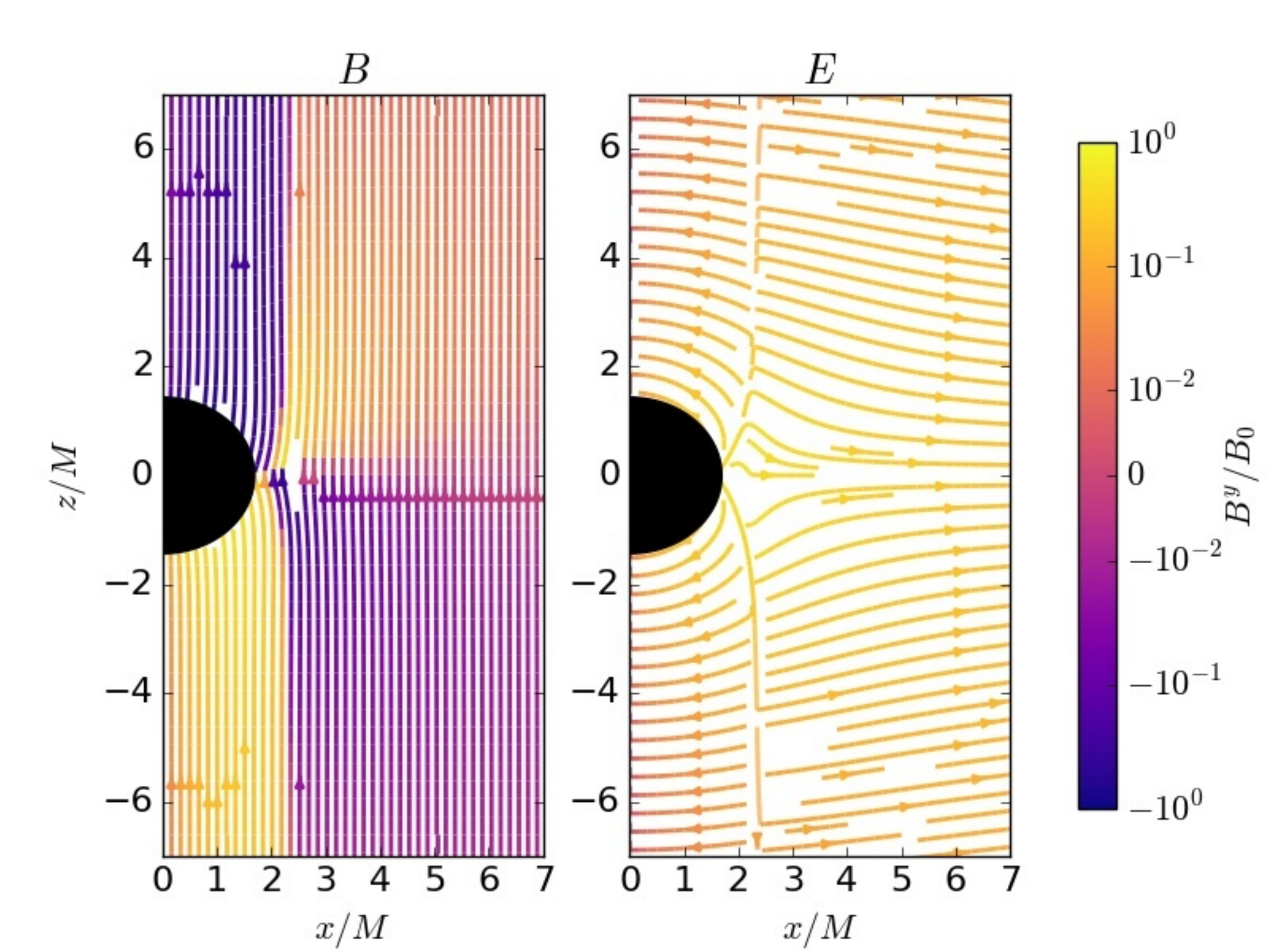}
\end{center}
\caption{
Streamlines of the magnetic (left) and electric (right) fields around an 
$a=0.9$ BH. The streamlines indicate the components in the $x$-$z$ plane 
(where the BH spin and the asymptotic magnetic field points in the $z$ direction)
    while the color indicates the out-of-plane ($y$) components of the fields.
\label{fig:stream}
}
\end{figure}

We start with an asymptotically uniform magnetic field and evolve
with FFE in the presence 
of a BH with aligned spin for times $\gtrsim 10^3 M$. For all cases we find that the
electromagnetic fields relax to a stationary, jetlike solution
which we study for various values of the BH spin $a$.
In Fig.~\ref{fig:stream}, we show an example of the configuration of 
the field lines for one such case, consistent with previous results~\cite{Komissarov:2004ms}.

\begin{figure}
\begin{center}
\includegraphics[width=\columnwidth,draft=false]{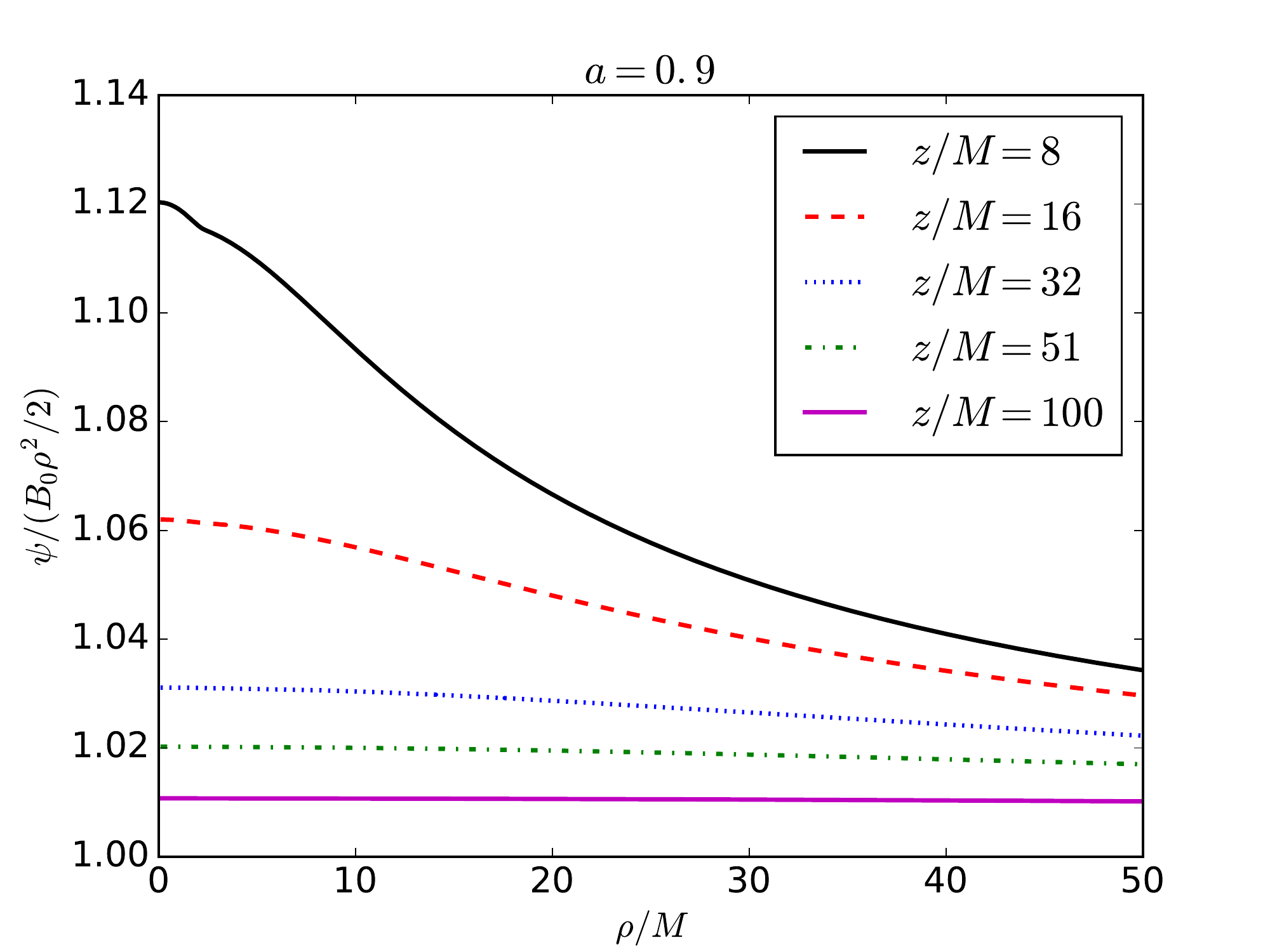}
\end{center}
\caption{
The magnetic potential $\psi$ on surfaces of fixed $z$ coordinate,
as a function of $\rho:=\sqrt{x^2+y^2}$.
As expected, $\psi$ approaches $\rho^2$ like $1/r$ at large distances.
This is shown for $a=0.9$, but the other cases are similar.
\label{fig:psiz}
}
\end{figure}

\begin{figure}
\begin{center}
\includegraphics[width=\columnwidth,draft=false]{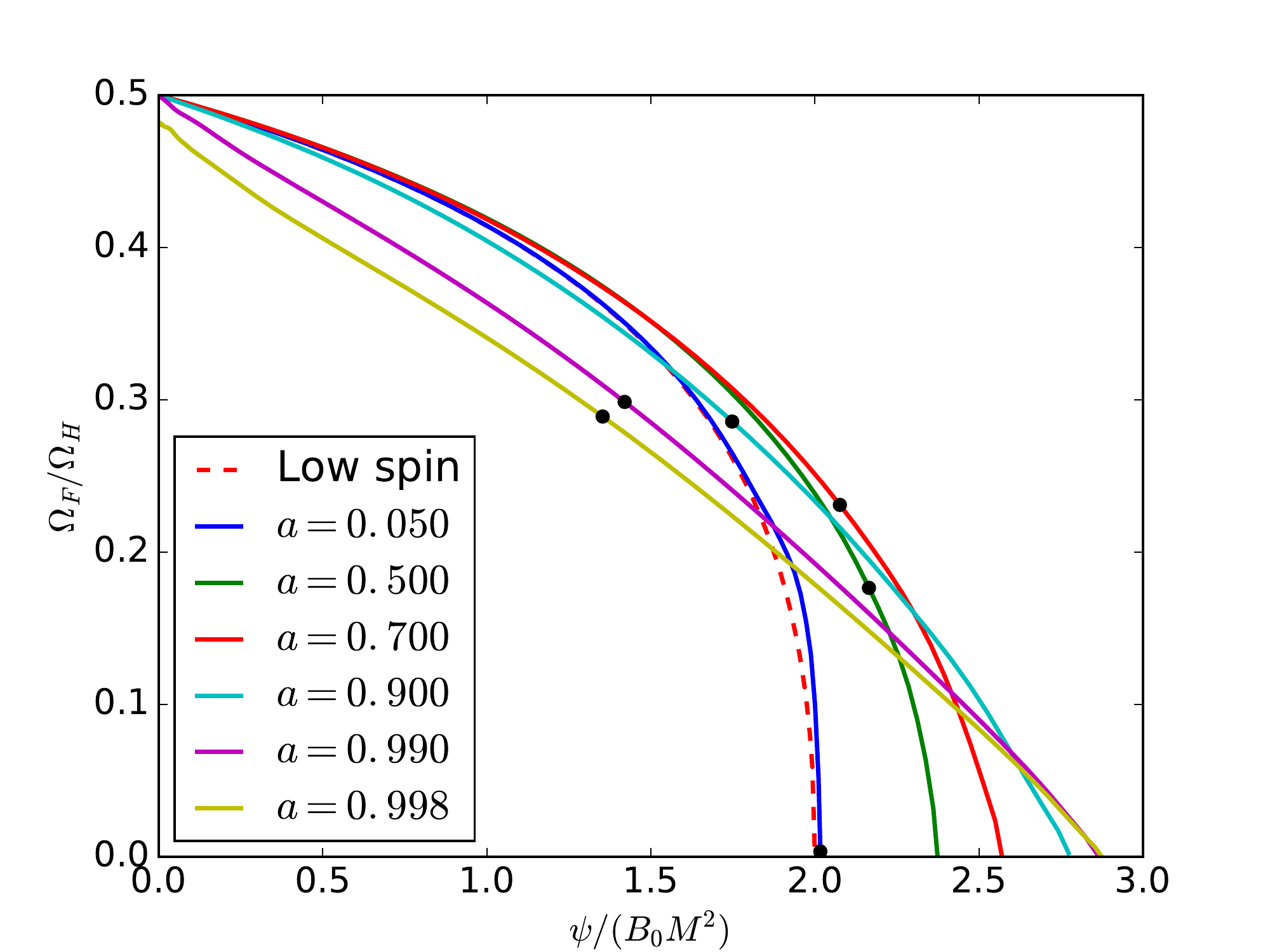}
\includegraphics[width=\columnwidth,draft=false]{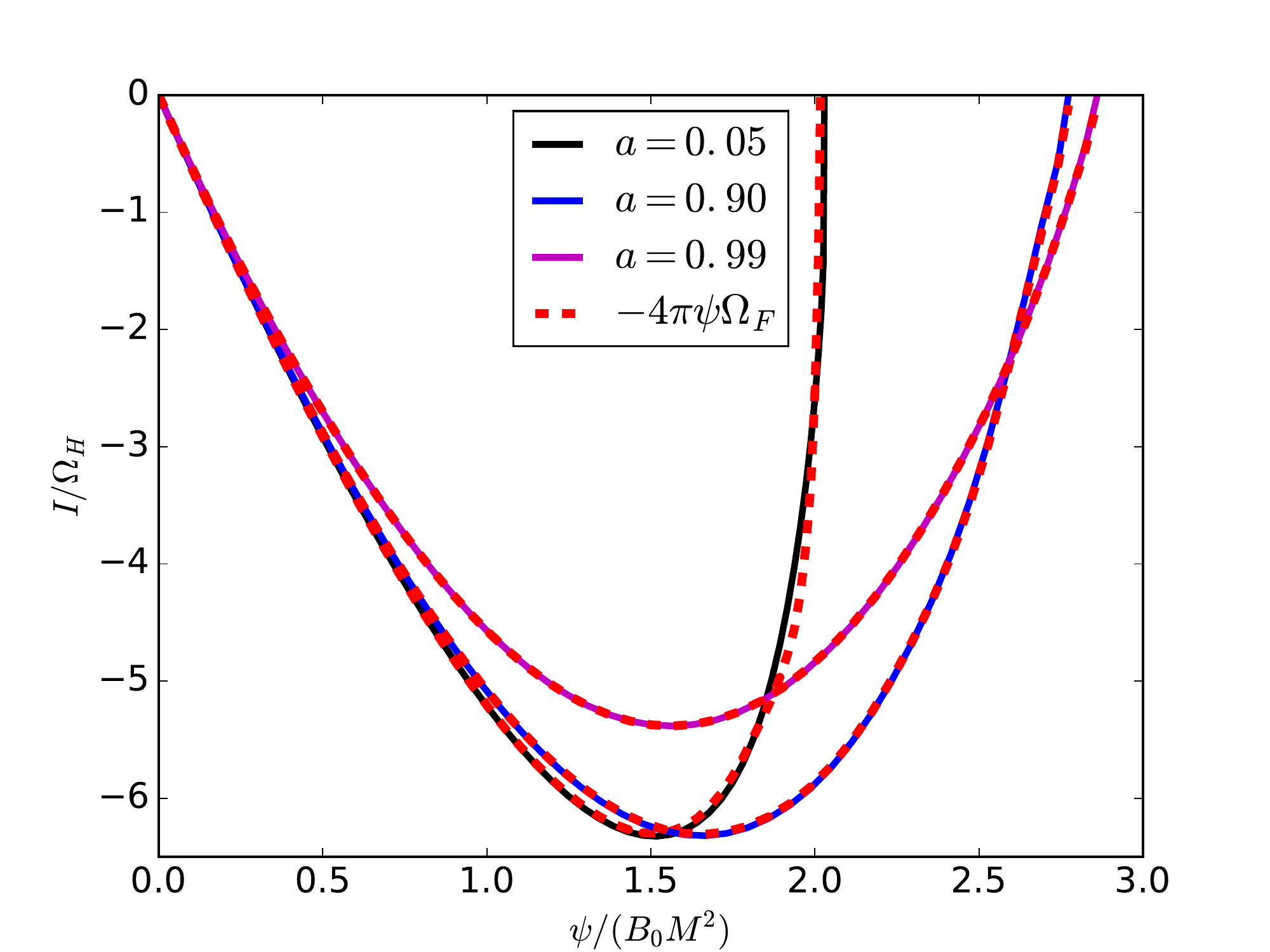}
\end{center}
\caption{
Top: The angular velocity of the field lines $\Omega_F$ as a function of 
$\psi$ for various spins. The black dots indicate the potential
of the last field line to cross the BH horizon.
The dashed red line is the low-spin approximation given by Eq.~\eqref{eqn:omega_psi}.
Bottom: The polar current $I$ as a function of $\psi$ for various BH 
spins. In all cases the dependence matches $I=-4\pi\Omega_F \psi$, indicated
by the dashed, red curves.
\label{fig:psi}
}
\end{figure}

For all values of BH spin, we find that $\psi$ asymptotes to $B_0\rho^2/2+\mathcal{O}(1/r)$
 at large distances. This is illustrated in Fig.~\ref{fig:psiz}.
For each value of $\psi$, we can measure the polar current 
$I$ and the angular velocity of the field lines $\Omega_F$.
We find that $\Omega_F$ is positive for the last field line
touching the BH horizon, and only vanishes for the last
field line entering the BH ergosphere. In between are
the field lines that hit the current sheet in the equatorial plane
of the ergosphere. This is illustrated in the top panel of Fig.~\ref{fig:psi}.
The bottom panel shows polar current, which always obeys the relationship
$I=-4\pi\Omega_F \psi$, and also vanishes for field lines that do not
enter the BH ergosphere.

\begin{figure}
\begin{center}
\includegraphics[width=\columnwidth,draft=false]{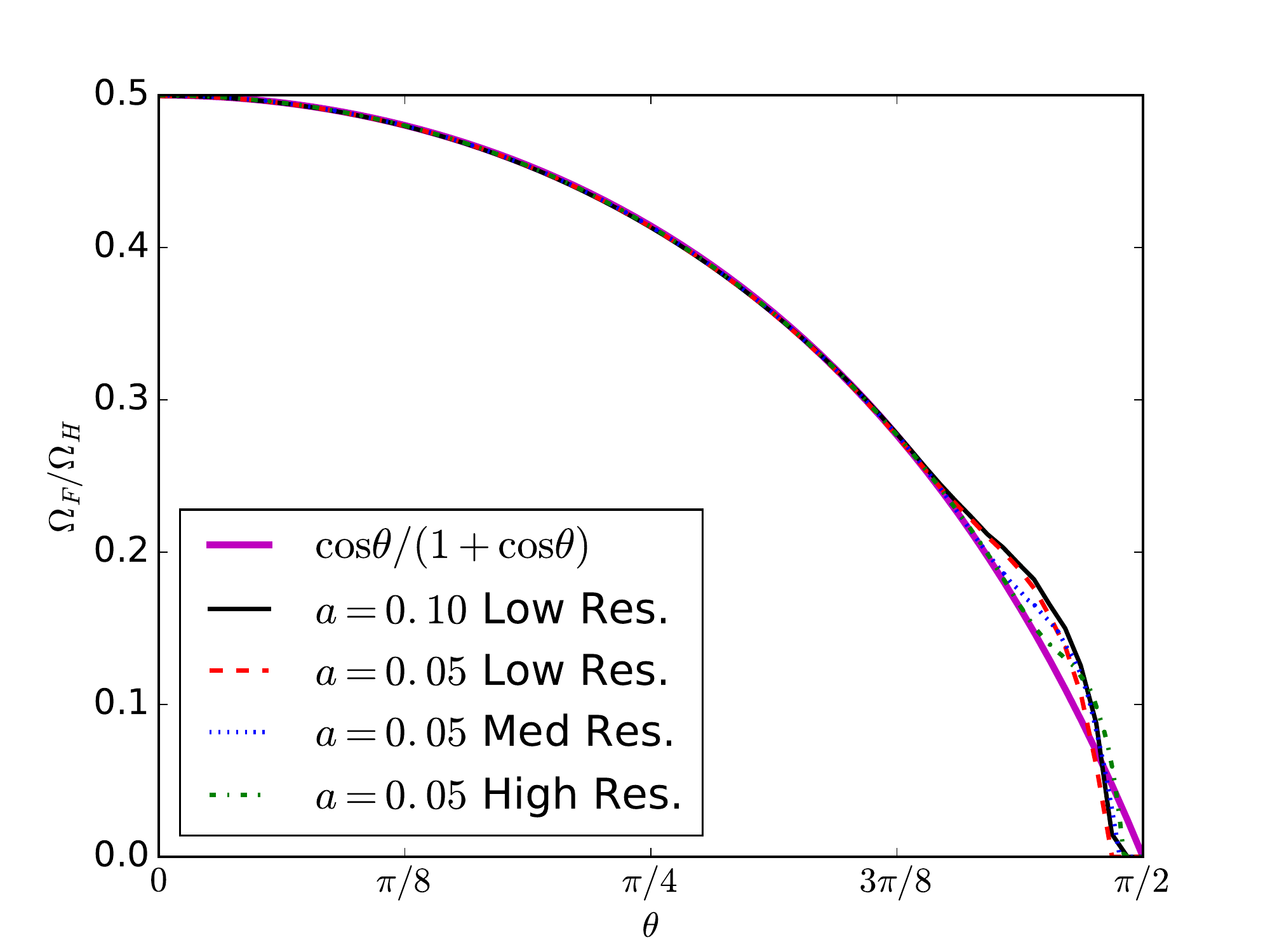}
\includegraphics[width=\columnwidth,draft=false]{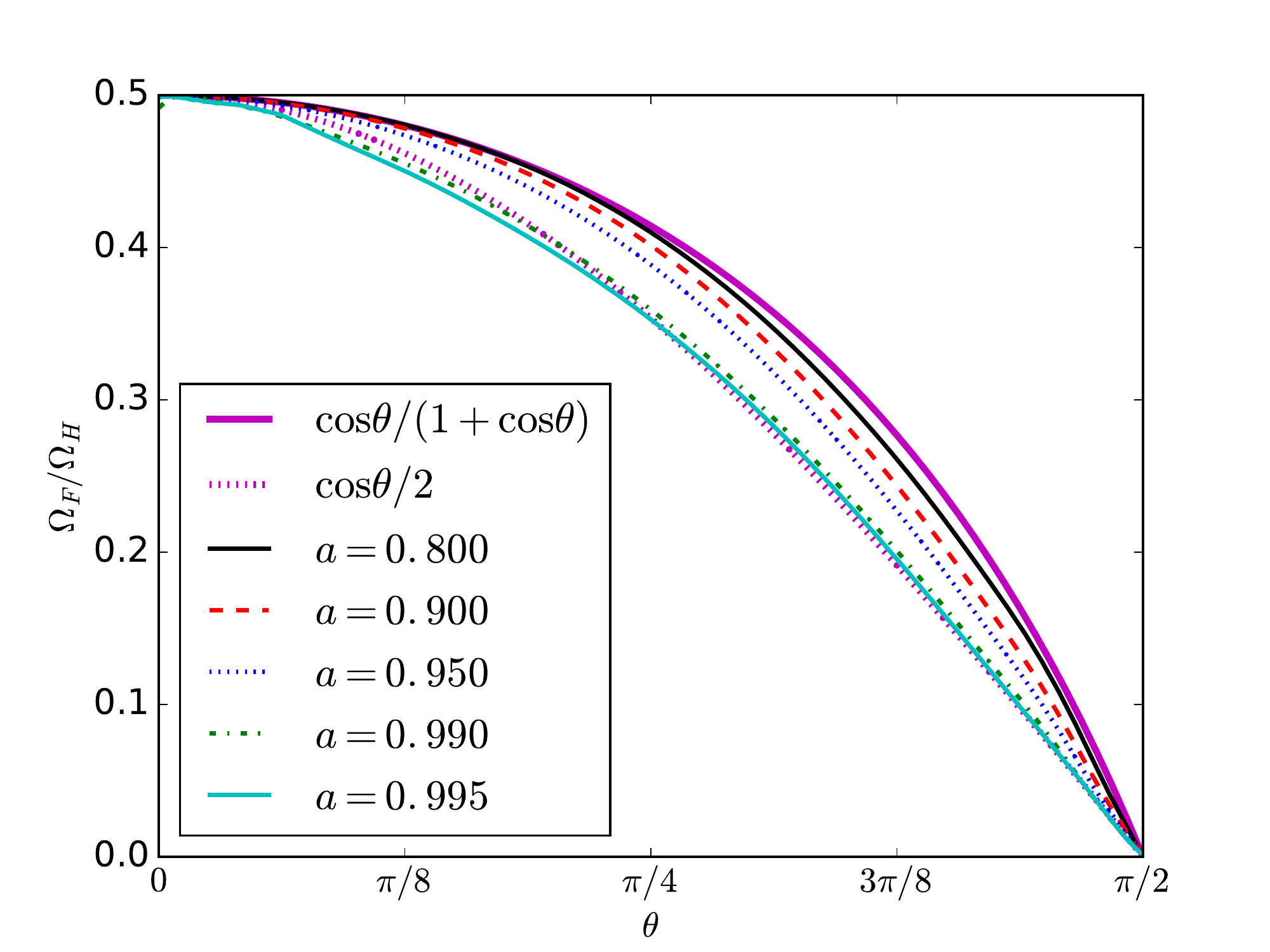}
\end{center}
\caption{
The rotational frequency of field lines $\Omega_F$ on the ergosphere,
as a function of Boyer-Lindquist coordinate $\theta$.
The top panel shows low spin cases, which approach the dependence 
    given by Eq.~\eqref{eqn:omega_theta}, modulo resolution dependent effects.
The bottom panel shows a range of spins. For high spins, the dependence seems closer
to $\cos\theta /2$ (indicated by the dotted magenta line). 
\label{fig:omegaf}
}
\end{figure}

We also show $\Omega_F$ on the boundary of the ergoregion as a
function of Boyer-Lindquist polar angle $\theta$ in Fig.~\ref{fig:omegaf}. For
small values of the BH spin (top panel), $\Omega_F$ does indeed obey
Eq.~\eqref{eqn:omega_theta}, with any differences shrinking with increased
numerical resolution. (We recall that in the low spin limit, the ergoregion
approaches the BH horizon.) This relation seems to approximately hold
even to larger values of $a\approx 0.8$ (bottom panel).  For near extremal BH
spins, $\Omega_F/\Omega_H$ has a shallower dependence that is closer to $\cos
\theta /2$.

\begin{figure}
\begin{center}
\includegraphics[width=\columnwidth,draft=false]{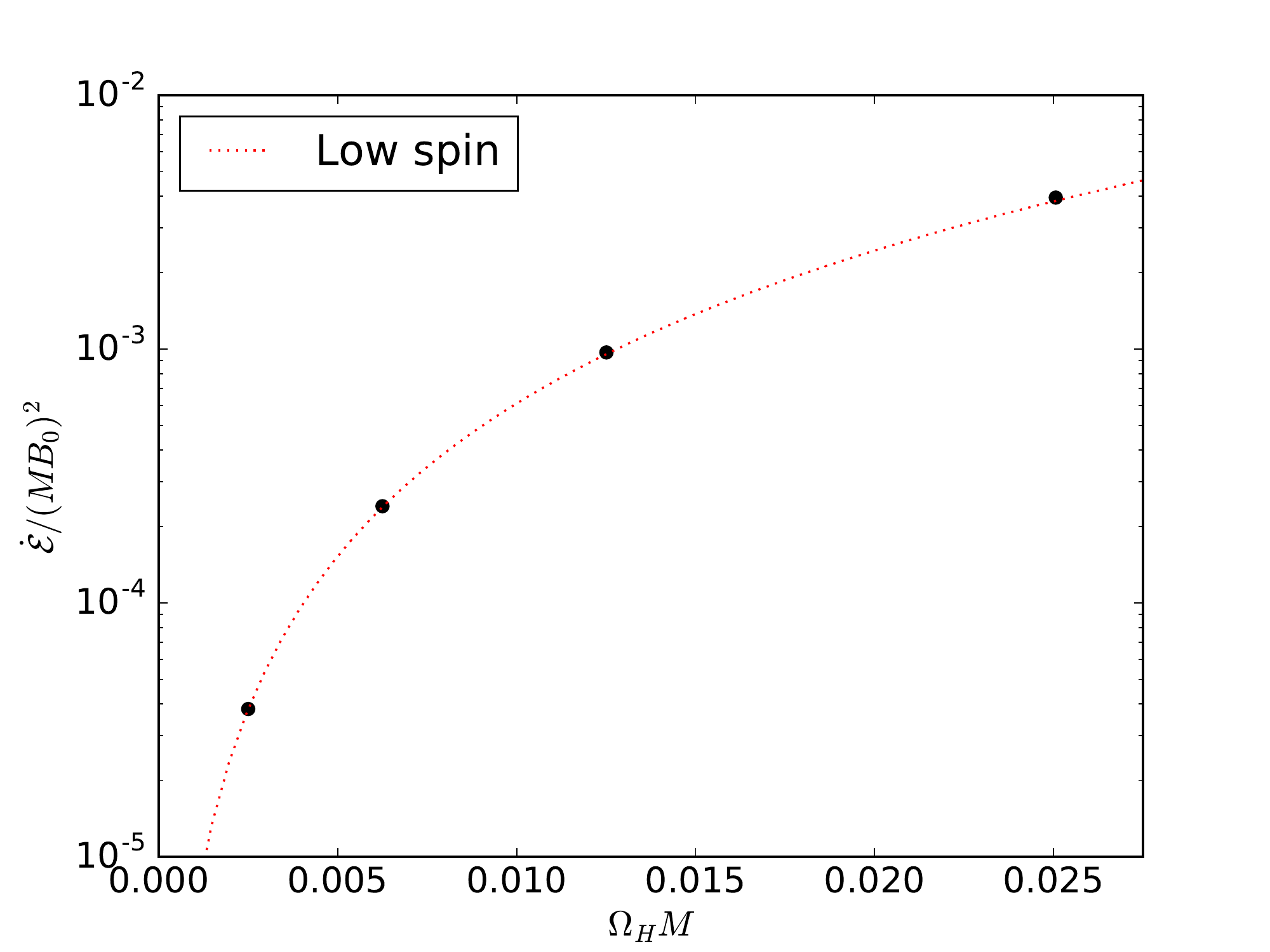}
\includegraphics[width=\columnwidth,draft=false]{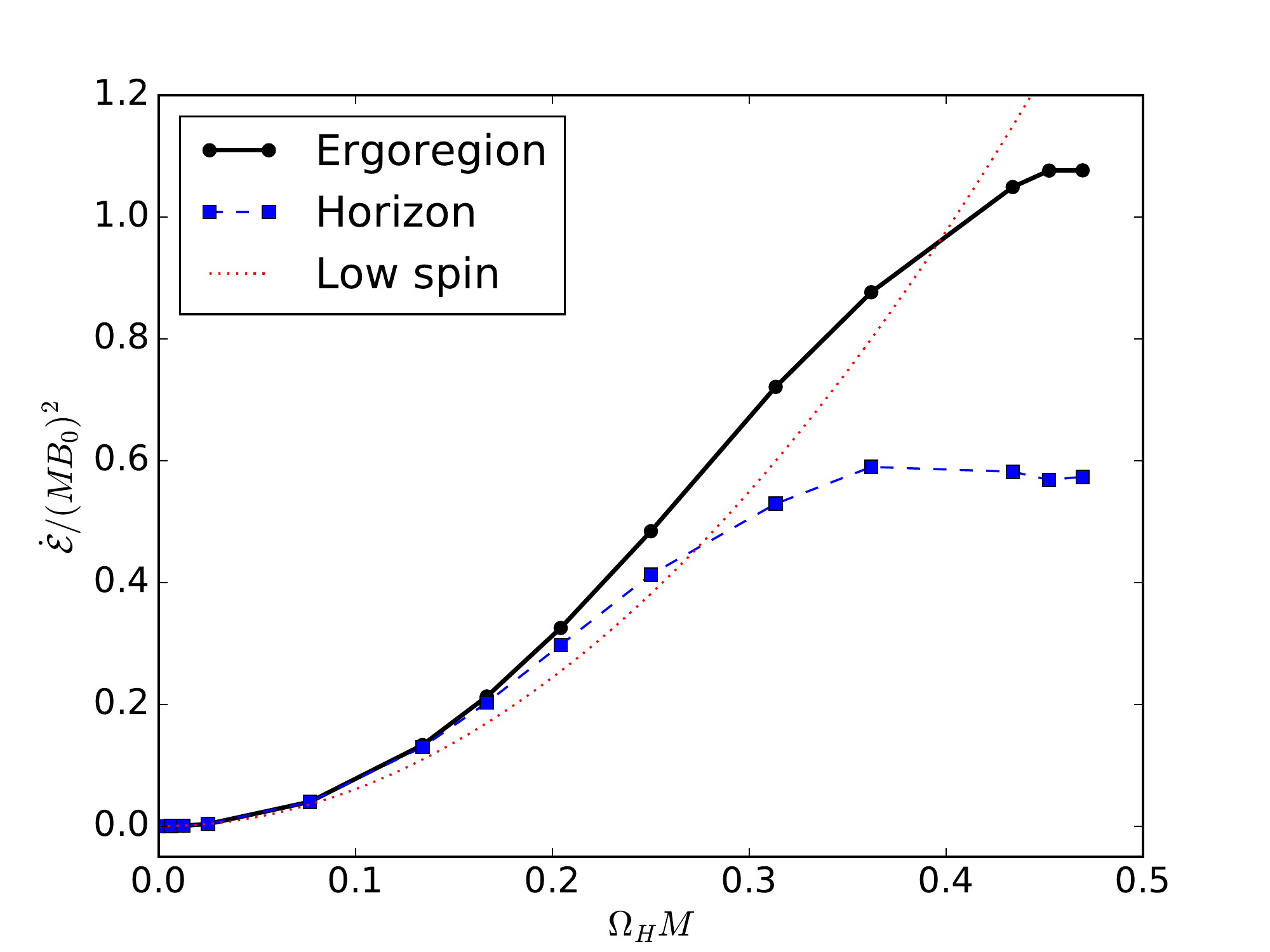}
\includegraphics[width=\columnwidth,draft=false]{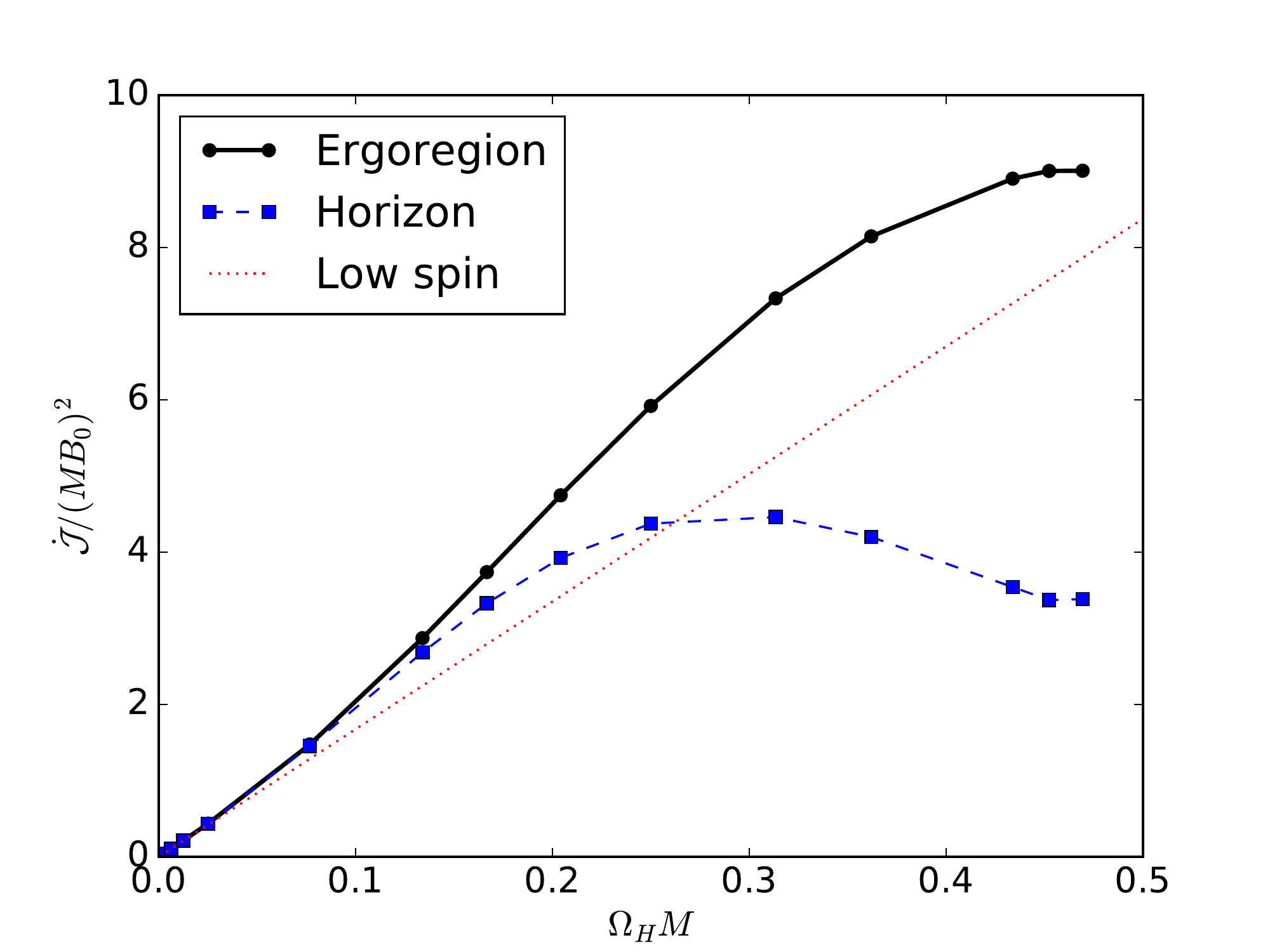}
\end{center}
\caption{
The flux of energy (top and middle panels) and angular momentum (bottom) through 
the BH horizon (blue squares) and the boundary of the ergoregion (black dots).
The difference between these two is due to the current sheet that
forms in the equatorial plane of the ergoregion.
    The dotted red curves indicate the low spin approximations given by Eqs.~\eqref{eqn:edot} and \eqref{eqn:jdot}. 
\label{fig:edot}
}
\end{figure}

We can also calculate the flux of energy and angular momentum coming from the
jet.  Since the force-free equations conserve energy and angular momentum in
axisymmetric, stationary spacetimes, for a stationary jet solution, one would
expect the flux of these quantities from the jet to be equal to the flux
through the BH horizon. However, one finds instead that the energy and
angular momentum flux through a surface at or outside the ergosphere is greater
than through the BH horizon.  The reason for this difference is the breakdown
of the force-free equations at the current sheet. The values of
$\dot{\mathcal{E}}$ and $\dot{\mathcal{J}}$ for these two different surfaces
are shown in Fig.~\ref{fig:edot}. For small spins, the role of the current
sheet is unimportant and these quantities are well approximated by the low-spin
expressions of Eqs.~\eqref{eqn:edot} and \eqref{eqn:jdot}, with negligible difference
between the horizon and the boundary of the ergosphere.
This is consistent with the $\dot{\mathcal{E}}\propto \Omega_H^2$ relation 
found in~\cite{Palenzuela:2010xn}.
For large spins, the
difference is quite pronounced, with roughly half the energy/angular momentum
flux coming from the current sheet.  We discuss how the
treatment of the breakdown of force-free at the current sheet affects this
result below.
We note in passing that, though we see no evidence of the BH Meissner 
effect in this setup---i.e. near extremal spin BHs do not expel magnetic field lines
from their horizon~\cite{Komissarov:2007rc}, the increase of $\dot{\mathcal{E}}$
and $\dot{\mathcal{J}}$ with $\Omega_H$ does appear to become small for near
extremal spin.

\subsection{Current sheet}
\begin{figure}
\begin{center}
    \includegraphics[width=\columnwidth,draft=false]{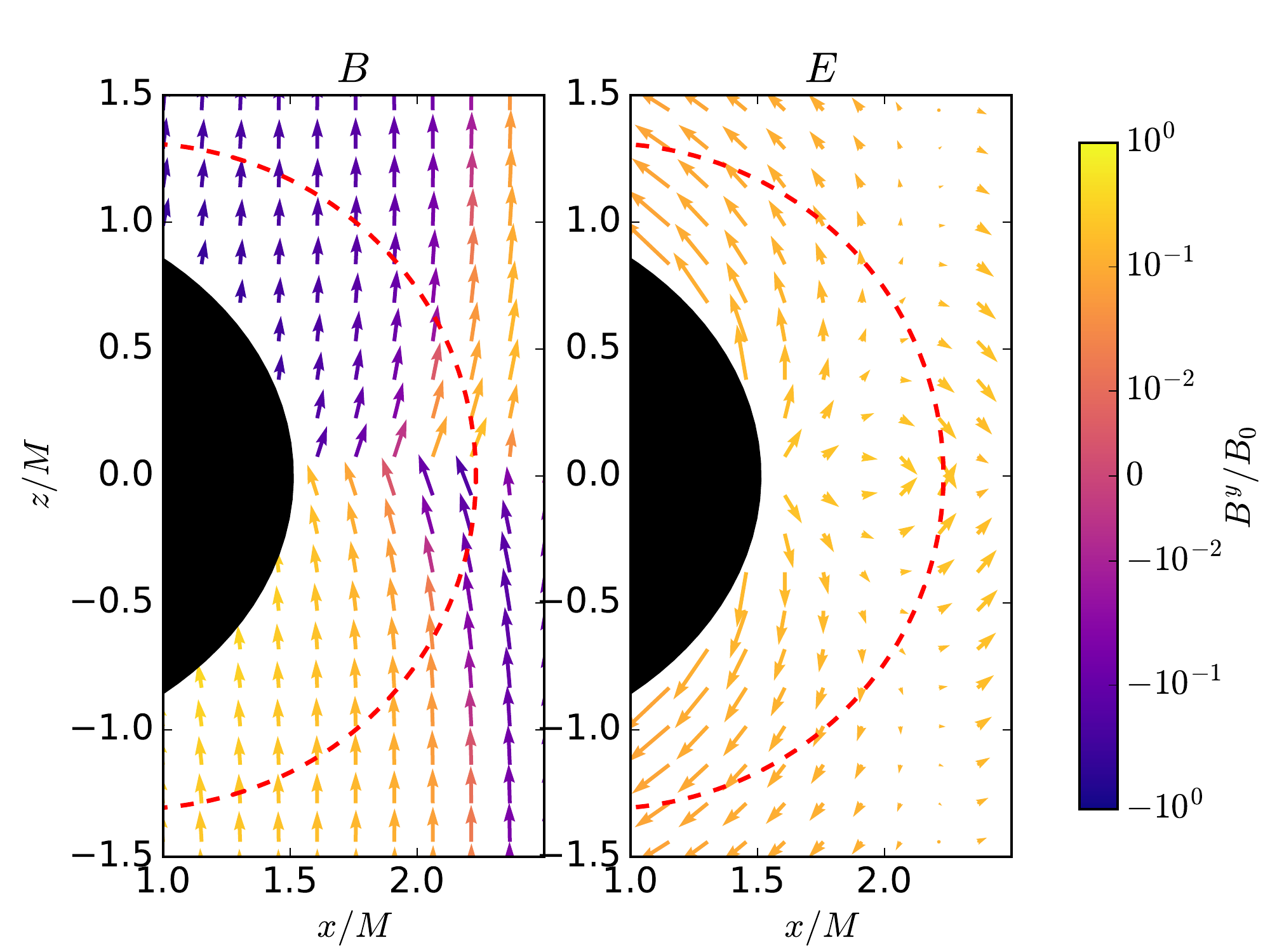}
    \includegraphics[width=\columnwidth,draft=false]{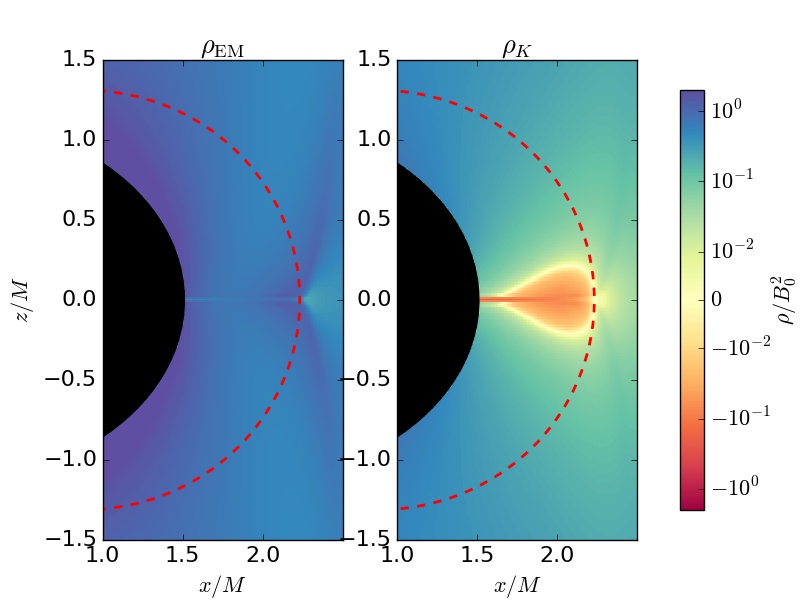}
\end{center}
\caption{
    Top: Quiver plot of the magnetic (left) and electric (right) fields in the
    neighborhood of the current sheet for a BH with $a=0.99$. The arrows
    indicate the components in the $x$-$z$ plane (where the BH spin and the
    asymptotic magnetic field points in the $z$ direction) while the color
    indicates the out-of-plane ($y$) components of the fields.  The dashed red
    line indicates the boundary of the ergosphere.  Bottom: Same as above, but
    showing the two measures of energy density defined in Sec.~\ref{ss:measure}.
\label{fig:quiver}
}
\end{figure}

Within the ergosphere, there is a discontinuity in the fields across the
equator: a current sheet.  The parallel components of the magnetic field and
the perpendicular components of the electric field flip sign across this
region, as can be seen in the top panel of Fig.~\ref{fig:quiver}.  (We find that if we enforce that these components are exactly zero on
the current sheet, the solution is more well-behaved in the neighborhood of the
current sheet, though elsewhere unchanged.) 
In the vicinity of the current sheet, the conserved energy density  
$\rho_K$ is negative (while the local
measure of energy density $\rho_{\rm EM}$ is
of course positive). This is illustrated in the bottom panel of Fig.~\ref{fig:quiver}.
Hence, locally dissipative processes occurring at the current sheet
where force-free breaks down
can lead to a positive contribution to the flux of energy from the jet.

\begin{figure}
\begin{center}
\includegraphics[width=\columnwidth,draft=false]{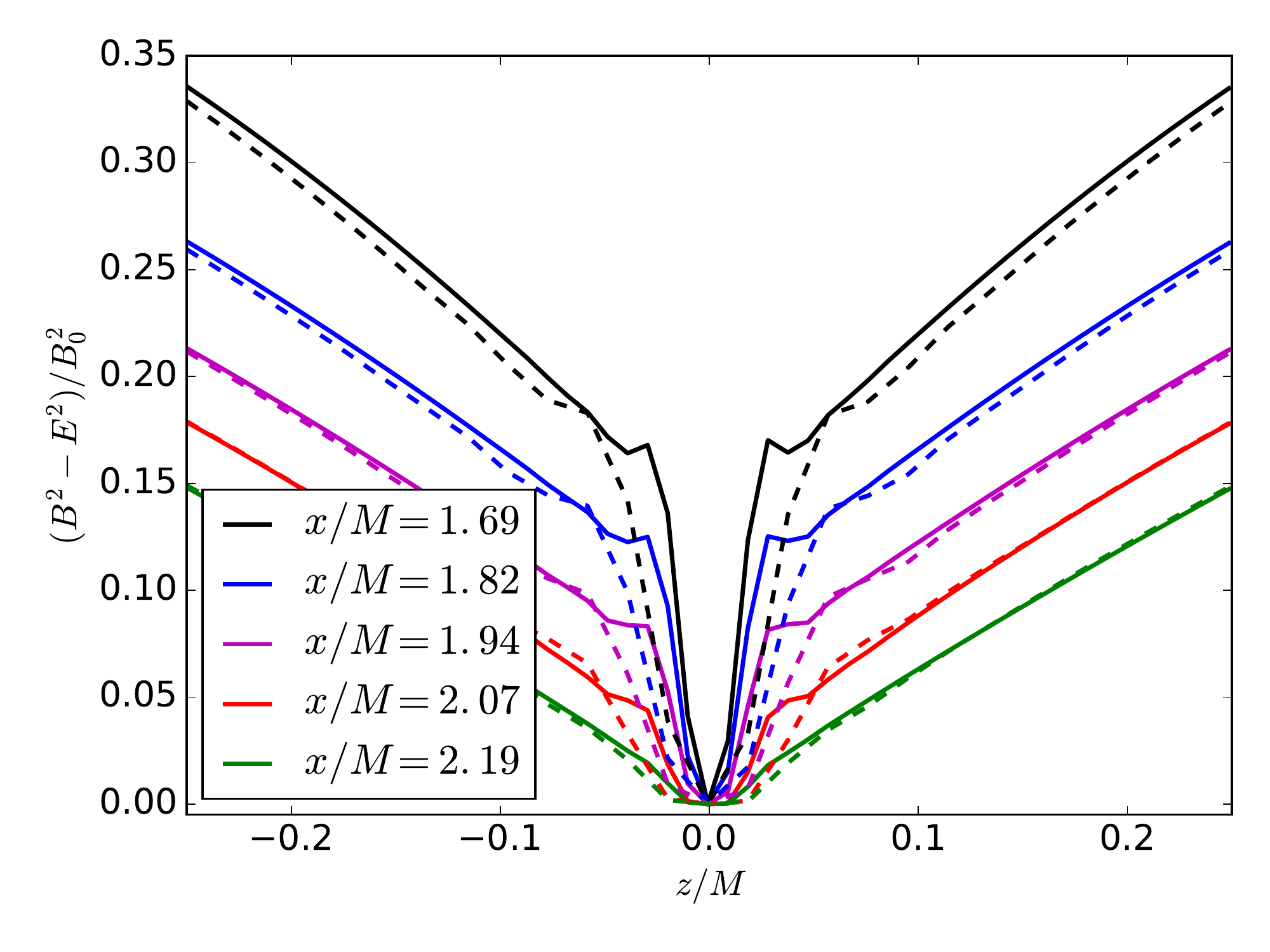}
\includegraphics[width=\columnwidth,draft=false]{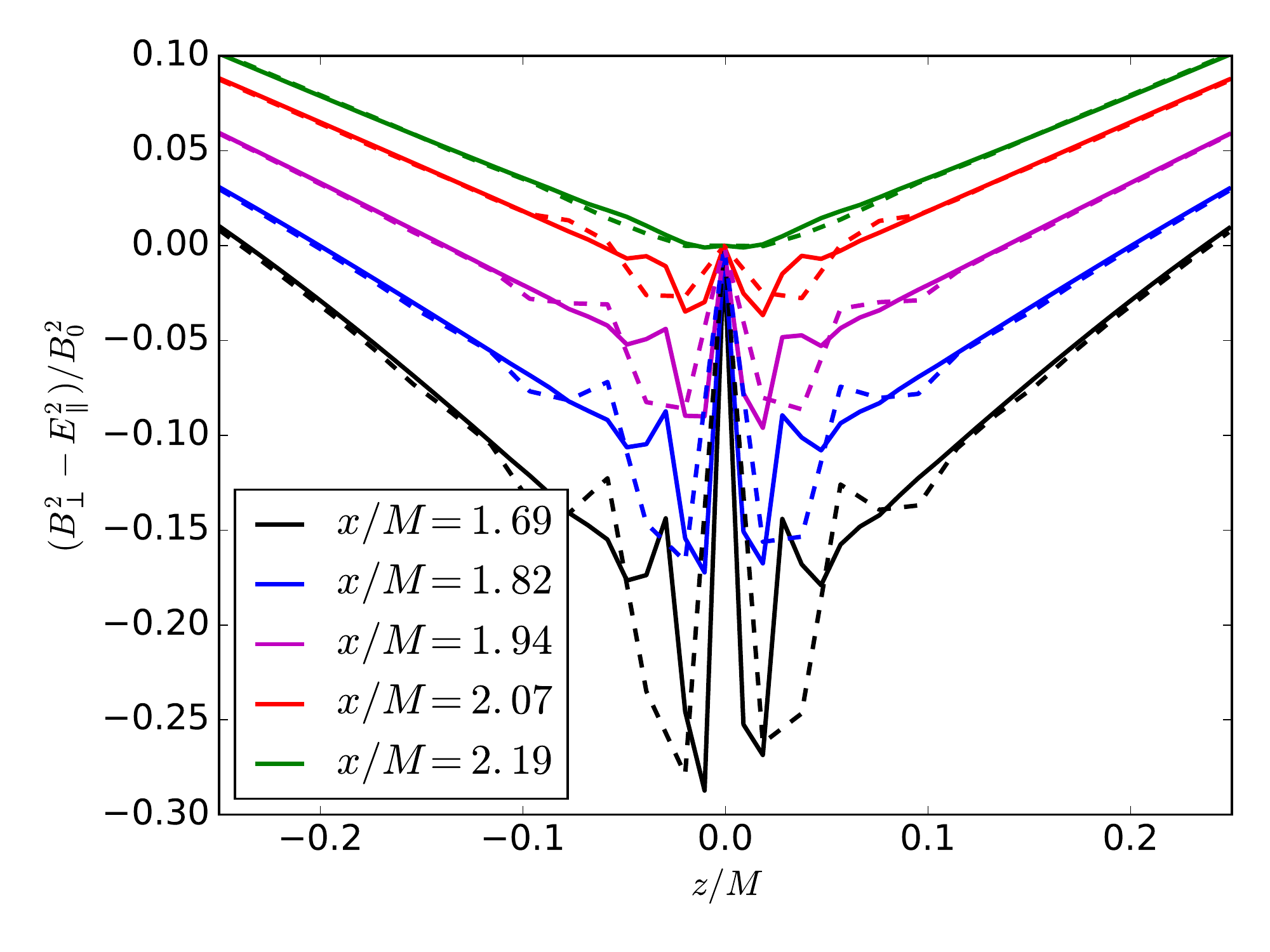}
\end{center}
\caption{
    Top: The value of $F^2/2=B^2-E^2$ along lines of constant $x$ passing
    through the current sheet at $z=0$ for a BH with $a=0.99$. The black line
    crosses $z=0$ sheet approximately at the BH horizon, while the green line
    crosses $z=0$ at the outer boundary of the ergosphere. The solid and dotted
    lines correspond to a higher and lower resolution, just to illustrate that
    the width of the discontinuous region is controlled by the resolution, but
    not the value being approached from above or below.  Bottom: Same as the
    top panel, but only showing the contribution from the components that do
    not jump across the current sheet.
\label{fig:fsq}
}
\end{figure}

The breakdown of force-free is
signaled by the quantity $F^2$ evolving towards a nonpositive value. This
occurs for these BH-jet solutions on the current sheet, though with the
prescription described in Sec.~\ref{ss:mag_dom}, we force $F^2=0$. However,
$F^2$ actually has a positive limiting value approaching the current sheet from
above or below, except at the edge of the ergosphere, where $F^2$ smoothly goes
to zero. This can be seen in Fig.~\ref{fig:fsq}. 

There are two contributions to $F^2$  
\beq
\frac{1}{2}F^2=B^2-E^2=(B_{\parallel}^2-E_{\perp}^2)+(B_{\perp}^2-E_{\parallel}^2).
\eeq
The
first term in parentheses is the contribution from the parallel components of the magnetic field
and the perpendicular components of the electric field, which is forced to
vanish when these components pass through zero in the current sheet. The second
contribution is from the field components that are continuous across the
current sheet: the perpendicular component of the magnetic field and the
parallel components of the electric field 
\footnote{
The covariant way to describe this would be to say that we can decompose the
field strength tensor into its pullback to the three-dimensional world volume of the
current sheet surface, and a perpendicular component, and that the condition
across the current sheet is that the jump in the former
vanishes~\cite{Gralla:2014yja}. Since here we have fixed coordinates where 
the current sheet is stationary, this is equivalent to our description in terms
of electric and magnetic fields.
}
.
As evident in the bottom panel of Fig.~\ref{fig:fsq},
this second contribution is negative approaching the current sheet, which
explains why magnetic dominance is lost at the current sheet when the first contribution
is zero. 

\begin{figure}
\begin{center}
\includegraphics[width=\columnwidth,draft=false]{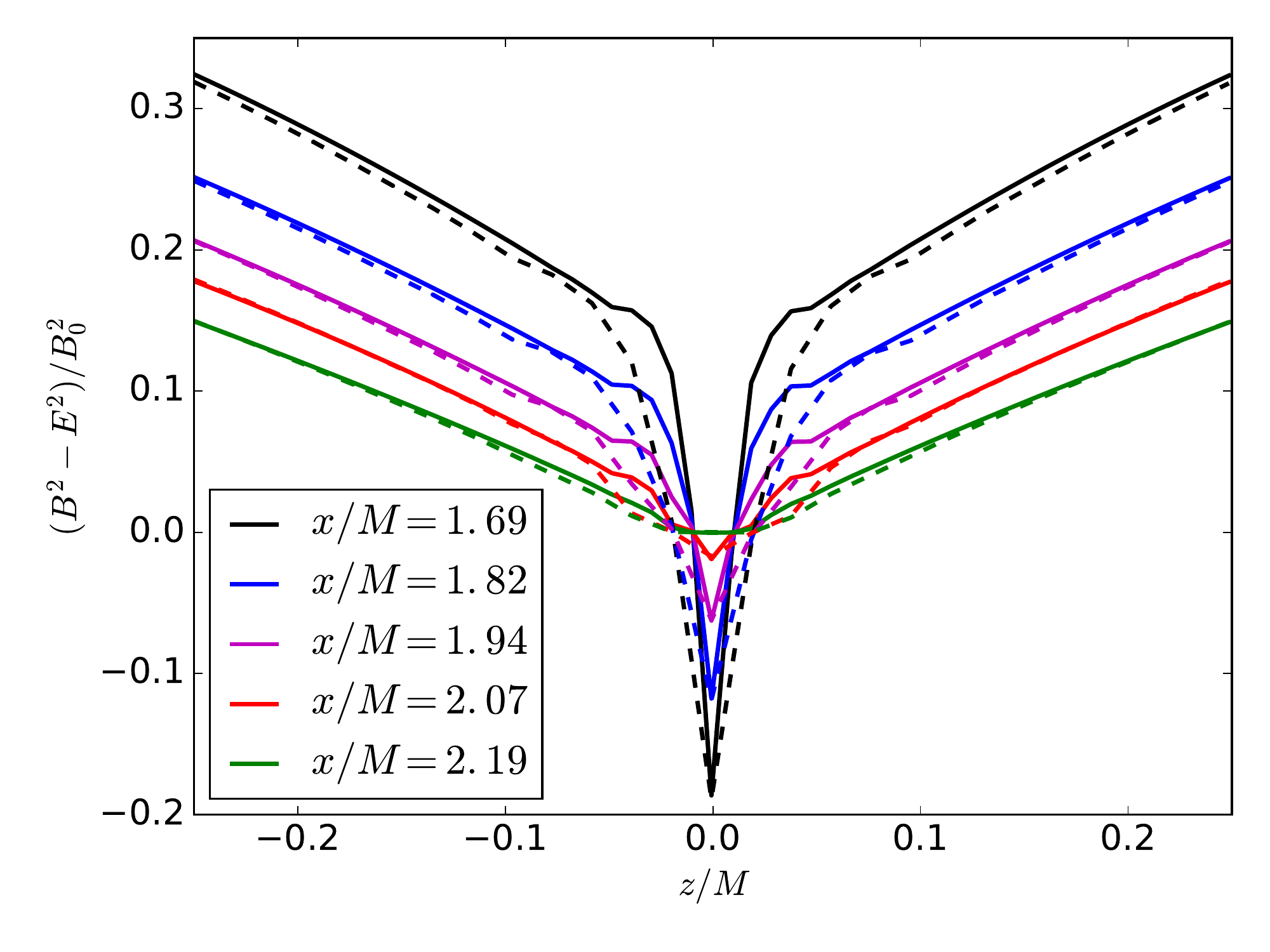}
\includegraphics[width=\columnwidth,draft=false]{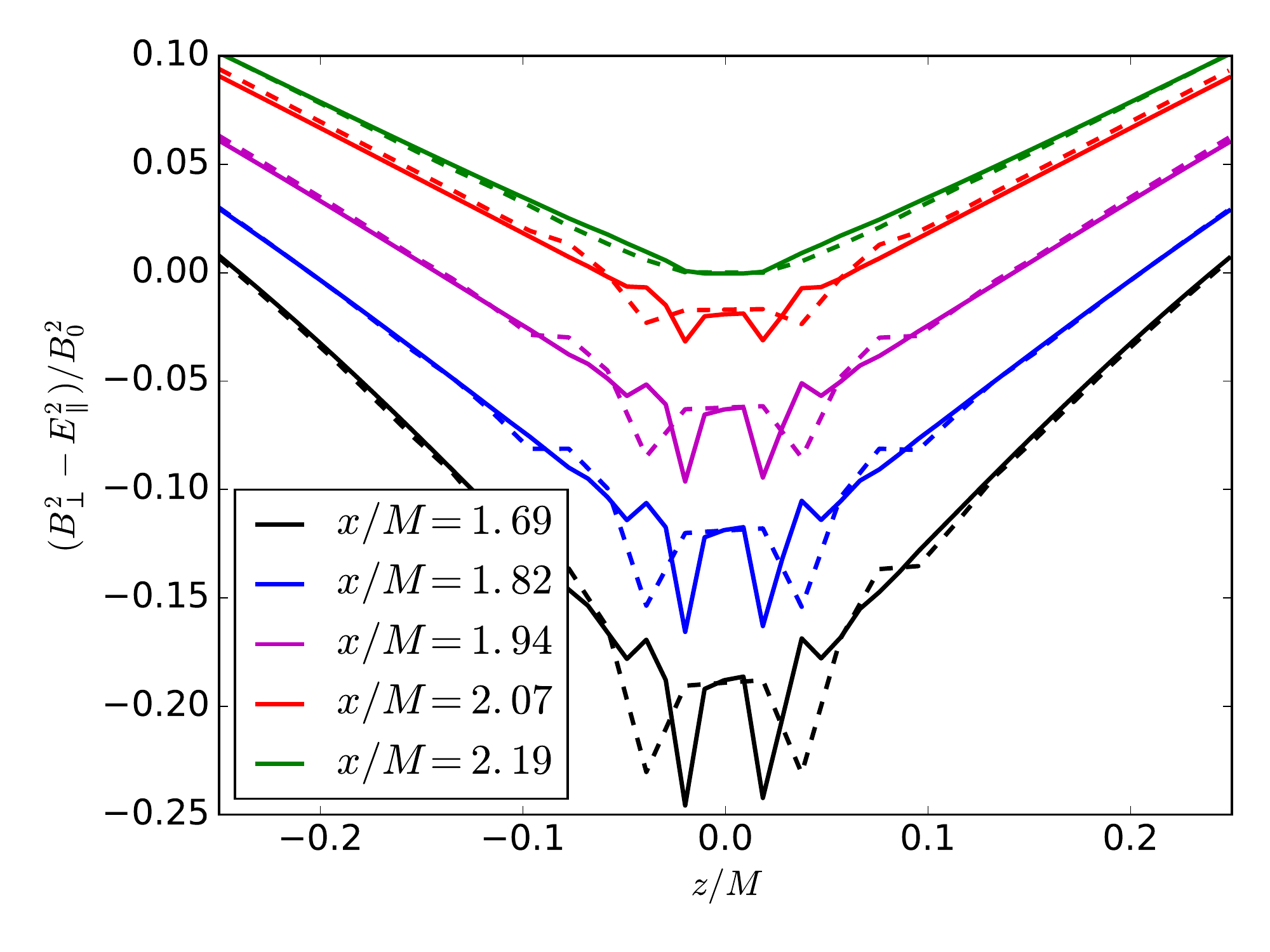}
\end{center}
\caption{
    The same as Fig.~\ref{fig:fsq}, but using a different treatment of the fields
    on the current sheet. Instead of rescaling the electric field to enforce 
    magnetic dominance, we set the parallel component of the magnetic field and 
    perpendicular components of the electric field by requiring continuity across
    the current sheet.
\label{fig:fsq_cs_bc}
}
\end{figure}

The prescription we apply to handle regions of $F^2\leq0$ is \emph{ad hoc}, and
ideally would by replaced by a microphysical description of the kinetic effects
of the plasma.  We can see from bottom panel of Fig.~\ref{fig:fsq} that this
condition forces the nominally ``continuous" field components (i.e. those that
are the same approaching the current sheet from above or below) to jump at $z=0$. 
In lieu of doing a kinetic calculation, we can study what happens if we apply a different condition
on the current sheet. In particular, in addition to setting $B_{\parallel}^2=E_{\perp}^2=0$
on the current sheet, we can enforce continuity in the other components of the electromagnetic
fields by setting them to be the average of the points at $z=\pm dz$. We apply this just to the
$z=0$ surface within the BH ergosphere, and find that $F^2>0$ everywhere else.
As shown in Fig.~\ref{fig:fsq_cs_bc}, with this prescription $B_{\perp}^2-E_{\parallel}^2$
no longer jumps to zero at $z=0$.
\footnote{There are still some numerical oscillations caused by the fact that we
use high-order finite differences that are not ideal for handling the discontinuities
in the fields across the current sheet, but these are restricted to a small region
controlled by the numerical resolution, and do not strongly affect the solution
elsewhere.
}
However, at the current sheet we now have that $E^2>B^2$. That would indicate
that within the current sheet there is a strong unscreened electric field with
$E^2$ reaching $\sim0.2B_0^2$ at the BH horizon (in the frame where the
magnetic field vanishes), for this case with $a=0.99$.  
Lower spin cases show similar behavior, though with smaller electric fields---e.g.
$E^2\sim0.02B_0^2$ at the BH horizon for $a=0.5$.
Applying this different condition at the current sheet
appears to have a small effect on the resulting solution elsewhere and,
e.g., the luminosity is essentially unchanged.
This condition was used for the results shown in Fig.~\ref{fig:quiver}.

\section{Discussion and Conclusion}
\label{sec:discuss}
In this work we have studied the force-free jet configurations obtained by evolving
the FFE equations until a stationary solution is approached, and found that in
the limit of low BH spin, these are well described by the solution derived in
Sec.~\ref{ss:low_spin}. This solution follows from demanding an outgoing
radiation condition: $I=-4\pi \Omega_F \psi$, which we find to hold for the jet
configurations that we obtain at all values of BH spin.  This differs
from~\cite{Yang:2015ata}, where specific solutions in the low spin limit were
identified based on maximizing the jet luminosity, or minimizing the energy
stored within the jet, but these solutions did not satisfy the above condition.
We find that in deriving this leading-order low spin solution, one may actually
ignore the effects of the current sheet.
The values of $\Omega_F$ we find via force-free evolution also differ somewhat
from those found using similar methods in~\cite{Yang:2015ata}, which may be
due to the higher resolution and/or longer relaxation times used here.

In~\cite{Nathanail:2014aua,Pan:2015imp,Pan:2017npg}, the uniform magnetic field
solution was studied by solving the Grad-Shafranov equations, which requires
the choice of boundary conditions along the nominal current sheet.
In~\cite{Nathanail:2014aua}, several arbitrary functions $\Omega_F(\psi)$ 
were considered in order to obtain a solution,
while in~\cite{Pan:2015imp,Pan:2017npg} the authors chose to make
the equator within the ergosphere a surface of equal potential, which means
that the last field line to enter the ergosphere intersects the BH horizon
horizontally. This is different from what is found here by evolving the FFE
equations.  Because of the breakdown of the force-free equations at the current
sheet, fields lines entering the ergosphere do not necessarily intersect the BH
horizon.  Obtaining similar solutions as found here by solving the
Grad-Shafranov equations presumably requires that a different boundary
condition be placed on the equator that captures the role of the current sheet,
though it is not obvious what condition to use.

Here we have also quantified how much of the flux of energy and angular
momentum coming from the jet is actually coming from the BH horizon, as opposed
to being due to the current sheet that forms on the equator within the
ergosphere. We have found that for rapidly spinning BHs, the latter contributes roughly as
much as the former.  This is counterintuitive as one typically
thinks of current sheets as being the site of the dissipation of
electromagnetic energy.  However, because it occurs within the ergosphere, it
is possible for a process that looks locally dissipative to correspond to a
gain of energy (or the annihilation of negative energy) as seen by a far-away observer
(see~\cite{1990ApJ...354..583P} for a related mechanism).

Of course, even if the force-free solution is giving the correct solution
elsewhere, it breaks down at the current sheet and can not describe how the
dissipated (positive or negative) energy goes into accelerating or heating
particles.  We have shown that the limiting values of the electromagnetic
fields approaching the current sheet obey magnetic dominance (i.e. $B^2>E^2$)
except at the equatorial boundary of the ergosphere (where $B^2=E^2$).
Furthermore, when one does not artificially impose magnetic dominance at the
current sheet, the limiting values approaching the current sheet suggest that
the fields should become electrically dominated within the current sheet.  This
strong, unscreened electric field could accelerate particles. Some of these
would fall into the BH horizon carrying negative energy as seen by a distant
observer, while others could escape to power high-energy radiation, \emph{\`a
la} the original particle Penrose process.  However, determining if and how
this occurs requires a kinetic calculation, e.g. using particle-in-cell
methods, which is something that we leave for future work. 

The results obtained here also shed light on those of~\cite{Ruiz:2012te}, where
it was shown that regular spacetimes with ergospheres can also power jets in
FFE.  Since such solutions also develop current sheets, they still have a site
for the dissipation of negative energy, even though there is no BH horizon.
For future work, it would be interesting to apply the methods used here to
study the problem of boosted BH(s) in a uniformly magnetized
plasma~\cite{Neilsen:2010ax,palenzuela2010dual,yang2016plasma}, in order to
understand how jets are powered in that case. 

\acknowledgments
We thank Luis Lehner, Zhen Pan, Mohamad Shalaby, and Jonathan Zrake for stimulating
discussions and comments on a draft of this work.  Simulations were run on the
Perseus Cluster at Princeton University, the Sherlock Cluster at Stanford
University, and the Comet Cluster at the San Diego  Supercomputer  Center
through XSEDE grant AST15003.  H.~Y.~acknowledges the support of the Natural
Sciences and Engineering Research Council of Canada. This research was
supported in part by Perimeter Institute for Theoretical Physics. Research at
Perimeter Institute is supported by the Government of Canada through the
Department of Innovation, Science and Economic Development Canada and by the
Province of Ontario through the Ministry of Research, Innovation and Science.  

\bibliographystyle{apsrev4-1.bst}
\bibliography{ref}
\end{document}